\documentclass[twocolumn,prl,aps,showpacs]{revtex4}
\usepackage{graphicx}

\begin{document}

\title{\bf Conformational Properties of an Adsorbed
Charged Polymer}

\author{Chi-Ho Cheng$^{1,2}$}
\email{phcch@phys.sinica.edu.tw}

\author{Pik-Yin Lai$^{1,3}$}
\affiliation{$^1$Department of Physics and Center for Complex
Systems, National Central University, Taiwan \\ $^2$Institute of
Physics, Academia Sinica, Taiwan \\
$^3$Physics Division, National Center for Theoretical Sciences,
Taiwan}

\date{\today}

\begin{abstract}
The behavior of a strongly charged polymer adsorbed on an
oppositely charged surface of low-dielectric constant is
formulated by the functional integral method. By separating the
translational, conformational, and fluctuational degrees of
freedom, the scaling behaviors for both the height of the polymer
and the thickness of the diffusion layer are determined. Unlike
the results predicted by scaling theory, we identified the
continuous crossover from the weak compression to the compression
regime. All the analytical results are found to be consistent with
Monte-Carlo simulations. Finally, an alternative (operational)
definition of a charged polymer adsorption is proposed.
\end{abstract}


\maketitle

\vspace{-5pt}

% \section{I. Introduction}

Charged polymer (polyelectrolyte) adsorption on charged surface
remains an interesting and important problem due to its influence
to material science \cite{decher}, colloidal science
\cite{napper}, and biological science \cite{grosberg}. Hard
substrates, and soft surfactant layers at interfaces can also be
charged, due to the dissociation of ionic groups on the surfaces.
Because the electrostatic force is strong and long-ranged, the
electrostatic interaction between a charged polymer and a charged
surface usually dominates over other non-electrostatic ones.

The problem of charged polymer adsorption on charged surface can
be studied by many approaches \cite{netz1}. By replacing the
counterion effect by the Debye-H\"{u}ckel potential within the
linear mean-field theory, one solve the Edwards equation
\cite{wiegel,muthukumar,varoqui1}. One can also solve both the
Edwards equation and the Poisson-Boltzmann equation
self-consistently \cite{varoqui2,borukhov,chatellier} at nonlinear
mean-field level in which the effective screening length near the
charged surface may not be equal to the bulk one. Scaling theory
were also applied to the problem \cite{dobrynin1}. Even more, the
effect of attractive image forces from high-dielectric substrate
\cite{dobrynin2,cheng1,cheng2}, and repulsive image forces from
low-dielectric substrate \cite{borisov,yamakov,netz2,messina} were
also investigated by analytical methods or Monte-Carlo (MC)
simulation.

However, the analytical approaches involving the Edwards equation
usually impose zero monomer density at the charged surface in
which the electrostatic boundary condition cannot be faithfully
respected. It is only for the case of charged polymer adsorption
on the high-dielectric substrate studied by Cheng {\it et al.}
\cite{cheng1} that the surface monomer density is properly
treated. The surface monomer density follows a linear relation
with surface charge density at Debye-H\"{u}ckel level. It
indicates that the charged polymer is fully compressed on the
high-dielectric substrate without any conformational change. For
the low-dielectric substrate,  due to the repulsive image forces,
the polymer is not necessarily compressed on the substrate.
Instead, the conformational degree of freedom plays an important
role on the adsorption behavior.

In this paper, we study the conformational properties of charged
polymer adsorbed on the low-dielectric substrate at
Debye-H\"{u}ckel level by both the functional integral methods and
MC simulation. It is found that the usual Edwards equation is no
longer valid to describe the non-Gaussian feature of polymer
conformation. A new formulation by the functional integral method
is proposed and compared with simulation results. Finally we give
an operational definition of charged polymer adsorption.

% \section{II. Model}
A charged polymer carrying positive charges is immersed in a
medium ($z>0$) of dielectric constant $\epsilon$. At $z=0$ there
is an impenetrable surface. Below the surface ($z<0$), there is
the substrate of low dielectric constant $\epsilon'<\epsilon$.
Just above the substrate, there is an uniform surface charge
density $\sigma<0$. The adsorbed charged polymer always stays
above the surface charge layer. Denote the charge on a polymer
segment $ds$ by $q ds$, the Hamiltonian is
%\begin{widetext}
\begin{eqnarray}
{\cal H} &=& \frac{1}{2} \int_0^N ds \int_0^N ds' \left(
\Gamma\frac{{\rm e}^{-\kappa|\vec r(s) - \vec r(s')|}}{|\vec r(s)
- \vec r(s')|} \right.
+ \left. \Gamma'(2-\delta_{s,s'}) \right. \nonumber \\
&& \left. \times \frac{{\rm e}^{-\kappa|\vec r(s) - \vec
r'(s')|}}{|\vec r(s) - \vec r'(s')|} \right)
% \nonumber \\
%&&
- h \int_0^N ds \kappa^{-1}{\rm e}^{-\kappa \vec r(s)\cdot\hat
z} \label{ham}
\end{eqnarray}
%\end{widetext}
where $s$ is the variable to parametrize the chain and
$\kappa^{-1}$ the Debye screening length. $\vec
r(s)=(x(s),y(s),z(s))$, $\vec r'(s')=(x(s'),y(s'),-z(s'))$ are the
positions of the monomers and their electrostatic images,
respectively.  $\Gamma = q^2/\epsilon$, $\Gamma' =
\Gamma(\epsilon-\epsilon')/(\epsilon+\epsilon')>0$, and $h=4\pi
q|\sigma|/(\epsilon'+\epsilon)>0$ are the coupling parameters
governing the strengths of Coulomb interactions among the monomers
themselves, between the polymer and its image, and between the
polymer and the charged surface, respectively. Note that the above
Hamiltonian is not exact even at Debye-H\"{u}ckel level. In
particular, the longitudinal interaction decays algebraically
rather than exponentially \cite{netz2}. However, the
conformational properties related to the adsorption behavior will
not be affected. We shall focus on the case of a charged polymer
adsorption in a low ionic strength medium.

The continuum Hamiltonian in Eq.(\ref{ham}) is discretized to
perform MC simulation. The continuous curve $\vec r(s)$ is
replaced by a chain of beads $\vec r_i$ ($i=1,\ldots,N$) with
hard-core excluded volume of finite radius $a$.
%Polymer lengths up to $N=120$ are employed in simulation.
The length and energy units are $2a$ and $q^2/2\epsilon a$,
respectively. Runs up to $10^9$ MC steps and up to $N=120$ are
performed.
%to achieve good statistics.

% \section{III. Analytical Methods}

%Similar to the analytical treatment of the high-dielectric case,
%We first divide the charged polymer degree of freedom into the
%transverse and longitudinal directions.
The partition function of the system is
\begin{eqnarray}
Z &=& \int{\cal D}[\vec r(s)] \exp[-\frac{3}{2a^2}\int_0^N ds
\left(\frac{\partial {\vec r(s)}}{\partial s}\right)^2 - \beta
{\cal H}]
\nonumber \\
&=& \int \prod_{i=1}^N d(\Delta {\vec
r_i})\exp[-\frac{3}{2a^2}\sum_{i=1}^N (\Delta \vec r_i)^2 - \beta
{\cal
H}] \nonumber \\
&=& \int \prod_{i=1}^N d(\Delta {\vec r_{\parallel i}})d(\Delta
z_i)\exp[-\frac{3}{2a^2}\sum_{i=1}^N ((\Delta \vec r_{\parallel
i})^2+(\Delta z_i)^2)]
\nonumber \\
&& \times \exp[- \beta {\cal H}]
\end{eqnarray}
where $\vec r_\parallel(s)=(x(s),y(s))$ is the $xy$-plane
projection of the curve $\vec r(s)$. While the charged polymer is
adsorbed, $|\Delta z_i|\ll |\Delta \vec r_{\parallel i}|$, and
note that $\vec r_\parallel(s)$ should describe a 2D polymer
conformation. Hence we approximate
\begin{eqnarray}
Z &\simeq& \int \prod_{i=1}^N d(\Delta {\vec r_{\parallel
i}})d(\Delta z_i) \exp[-\frac{1}{a^2}\sum_{i=1}^N (\Delta \vec
r_{\parallel i})^2]
\nonumber \\
&&\times \exp[-\frac{1}{2a^2}\sum_{i=1}^N (\Delta \vec
r_{\parallel i})^2+(\Delta
z_i)^2] \exp[-\beta {\cal H}] \nonumber \\
&=& \int{\cal D}[\vec r_\parallel(s),\vec r_\perp(s)
]\exp[-\frac{1}{a^2}\int_0^N ds \left(\frac{\partial \vec
r_\parallel(s)}{\partial s}\right)^2 \nonumber \\
&& - \frac{1}{2a^2}\int_0^N ds \left(\frac{\partial \vec
r_\perp(s)}{\partial s}\right)^2 -\beta {\cal H}]
\end{eqnarray}
where $\vec r_\perp(s)=(|\vec r_\parallel(s)|,z(s))$ is the
side-view of $\vec r(s)$ along the curve $\vec r_\parallel(s)$.
Note that the coefficients of the entropy terms of $\vec
r_\parallel(s)$ and $\vec r_\perp(s)$ are $-1/a^2$ and $-1/2a^2$,
respectively, which are different from that of $\vec r(s)$,
$-3/2a^2$.

For the case of charged polymer adsorption, the self-electrostatic
interaction takes almost no effect in $\vec r_\perp(s)$ since
$|\vec r(s) - \vec r(s')| \simeq |\vec r_\parallel(s)-\vec
r_\parallel(s')|$. The repulsion from the images of the monomers
can be effectively approximated by the interaction between each
monomer and its image only. The residual repulsion is absorbed by
renormalizing $\Gamma'$. Then the partition function becomes
%\begin{widetext}
\begin{eqnarray} \label{part}
Z &\simeq& \int{\cal D}[\vec
r_\parallel(s)]\exp[-\frac{1}{a^2}\int_0^N ds \left(\frac{\partial
\vec r_\parallel(s)}{\partial s}\right)^2 \nonumber \\
&& - \frac{\beta\Gamma}{2}\int_0^Nds\int_0^Nds' \frac{{\rm
e}^{-\kappa|\vec r_\parallel(s) - \vec r_\parallel(s')|}}{|\vec
r_\parallel(s) - \vec r_\parallel(s')|} ]
\nonumber \\
&&\times \int{\cal D}[\vec r_\perp(s)] \exp[\int_0^N ds \{-
\frac{1}{2a^2}\left(\frac{\partial \vec r_\perp(s)}{\partial
s}\right)^2 \nonumber \\
&& - \frac{\beta\Gamma'}{4}
 \frac{{\rm e}^{-2\kappa \vec r_\perp(s)\cdot\hat z}}{\vec r_\perp(s)\cdot\hat z}
 + \beta h \kappa^{-1} {\rm e}^{-\kappa \vec r_\perp(s)\cdot\hat z} \}]
\end{eqnarray}
%\end{widetext}
The system is decoupled into two independent degrees of freedom,
$\vec r_\parallel(s)$ and $\vec r_\perp(s)$.
%$\vec r_\parallel(s)$ represents the self-electrostatic interaction in 2d.
% Its conformation is similar to that of 3d.
Since the above functional integral with respect to $\vec
r_\parallel(s)$ does not affect the adsorption behavior, we
investigate only the conformational properties of $\vec
r_\perp(s)$ in the following.

%%%%%%%%%%%%%%%%%%%%%%%%%%%%%%%%%%%%%%%%%%%%%%%%%%%%%%%%%%%%%%%%%%%%%%%%%%%%%%%%%%%%%%
\vspace{15pt}
\begin{figure}[tbh]
\begin{center}
\includegraphics[width=3in]{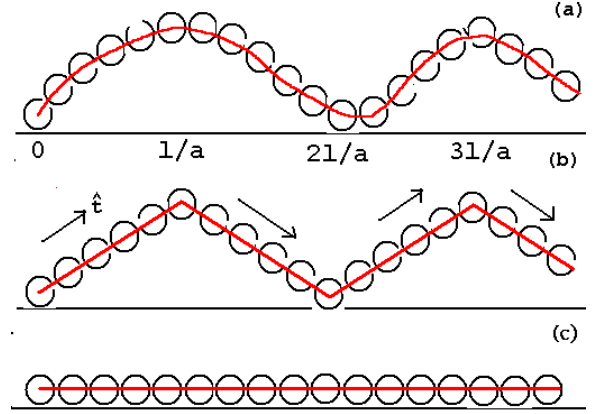}
\end{center}
%\vspace{-5pt}
 \caption{Schematic diagram for the conformation of an adsorbed charged polymer.
 The degrees of freedom (entropies) of the polymer consists of
 three parts, the translation ($\vec r_{\perp \rm c}$), the
 conformation ($\vec t(s)$), and the local fluctuation ($\delta \vec r_\perp(s)$).
 The solid line represents the polymer orientation. The local
 fluctuation lies within the blobs.
 (a) The weakly compressed polymer (onset of adsorption) conformation in general.
 (b) The weakly compressed polymer conformation in our analytical approximation.
 (c) The compressed polymer (adsorption) conformation in which the conformational degree of
     freedom $\vec t(s)$ vanishes.
 }
 \label{conf.eps}
 \vspace{-5pt}
\end{figure}

%%%%%%%%%%%%%%%%%%%%%%%%%%%%%%%%%%%%%%%%%%%%%%%%%%%%%%%%%%%%%%%%%%%%%%%%%%%%%%%%%%%%%%%

Because of the repulsive image force from the low-dielectric
substrate, the charged polymer may be at weak compression or
compression in which their schematic diagrams are shown in
Fig.\ref{conf.eps}a and \ref{conf.eps}c, respectively. The
terminology of weak compression (onset of adsorption) and
compression (adsorption) are borrowed from Borisov {\it et al.}
\cite{borisov,yamakov} for grafted polymers.

In order to distinguish between the weak compression and the
compression in our formulation, and note that a slowly varying
orientation of polymer conformation under weak compression, we
decompose
\begin{equation} \label{new_rs}
{\vec r_\perp(s)}={\vec r}_{\perp \rm c}+{\vec t(s)}+{\delta \vec
r_\perp(s)}
\end{equation}
where ${\vec r}_{\perp \rm c}=\frac{1}{N}\int_0^N ds {\vec
r_\perp(s)}$ is the position of the center of mass, and ${\vec
t(s)}$ is the orientation vector of the charged polymer. We also
restrict
\begin{equation} \label{new rs1}
\vec t(s)\cdot \delta \vec r_\perp(s)=0
\end{equation}
so that $\delta \vec r_\perp(s)$ represents the local fluctutation
along $\vec t(s)$.

The adsorbed polymer is now characterized by translational ($\vec
r_{\perp \rm c}$), conformational ($\vec t(s)$), and local
fluctutational ($\delta \vec r_\perp(s)$) degrees of freedom.
Under the compression regime, $\vec t(s)$ vanishes.

In general, it is hard to compute the effect from $\vec t(s)$. For
simplicity but still capturing the qualitative picture of the weak
compression as shown in Fig.\ref{conf.eps}a, we further make an
approximation that
\begin{eqnarray} \label{new_t}
{\vec t(s)}\cdot{\hat z} =\left\{\begin{array}{cc}
(2as/l-1) \vec r_{\perp \rm c}\cdot\hat z, \ &  0<s<l/a  \\
-(3-2as/l) \vec r_{\perp \rm c}\cdot\hat z, \  & l/a<s<2l/a
\end{array}\right.
\end{eqnarray}
and repeat for a period of $2l/a$. Its schematic diagram is shown
in Fig.\ref{conf.eps}b. Substituting
Eqs.(\ref{new_rs})-(\ref{new_t}) into Eq.(\ref{part}), and at
low-salt limit, we get
%\begin{widetext}
\begin{eqnarray}
Z &=&{\cal N}^{-1}\int d \vec r_{\perp \rm c}
 \exp [N\beta h {\vec r}_{\perp \rm c}\cdot\hat z]
\int{\cal D}[\delta \vec r_\perp(s)] \nonumber \\
&& \exp [\int_0^N ds \{
 -\frac{1}{2a^2}\left(\frac{\partial{\delta \vec r_\perp(s)}}{\partial{s}}\right)^2
 -\beta h {\delta \vec r_\perp(s)\cdot \hat z}
  \}] \nonumber \\
&&\times \int{\cal D}[\vec
t(s)]\exp[-\frac{\beta\Gamma'}{4}\int_0^N \frac{ds}{{\vec
r_\perp(s)}\cdot\hat z}]
\end{eqnarray}
%\end{widetext}
where the integral of $\vec r_\parallel(s)$ is absorbed into the
normalization constant $\cal N$. Expand the following integral
around small $\delta \vec r_\perp(s)$ up to quadratic order,
\begin{eqnarray}
\int_0^N \frac{ds}{{\vec r_\perp(s)}\cdot\hat z}
 = \frac{N}{l |\hat t\cdot\hat z|}
 \log
 \frac{2\vec r_{\perp \rm c}\cdot \hat z+l|\hat t\cdot\hat z|}
      {2\vec r_{\perp \rm c}\cdot \hat z-l|\hat t\cdot\hat z|}
\nonumber \\
 -\int_0^N ds \{\frac{\delta \vec r_\perp(s)\cdot \hat z}{(\vec
r_{\perp \rm c})^2} -  \frac{(\delta \vec r_\perp(s)\cdot \hat
z)^2}{(\vec r_{\perp \rm c})^3}\}
\end{eqnarray}
and then integrate out the variable $\vec t(s)$ under the
condition that $|\hat t\cdot\hat z|\ll 1$, the partition function
becomes
%\begin{widetext}
\begin{eqnarray} \label{part2}
Z &=& {\cal N}^{-1} \int_0^\infty dz_{\rm c} z_{\rm c}
\exp[-N\beta (h z_{\rm c}+\frac{\Gamma'}{4 z_{\rm c}})] \nonumber \\
&&\times \int{\cal D}[\delta z(s)]
 \exp [\int_0^N ds \{
 -\frac{1}{2a^2}\left(\frac{\partial{\delta
 z(s)}}{\partial{s}}\right)^2 \nonumber \\
&& -\beta( h -\frac{\Gamma'}{4z_{\rm c}^2})\delta z(s) -
\frac{\beta\Gamma'}{4z_{\rm c}^3}(\delta z(s))^2
  \}]
\end{eqnarray}
%\end{widetext}
Note that $l$ is related to $\vec t(s)$ via Eq.(\ref{new_t}), and
will be integrated out inside the functional integral of $\vec
t(s)$.

%%%%%%%%%%%%%%%%%%%%%%%%%%%%%%%%%%%%%%%%%%%%%%%%%%%%%%%%%%%%%%%%%%%%%%%%%%%%%%%%%%%%%%%
%\vspace{5pt}
\begin{figure}[tbh]
\begin{center}
\includegraphics[width=3in]{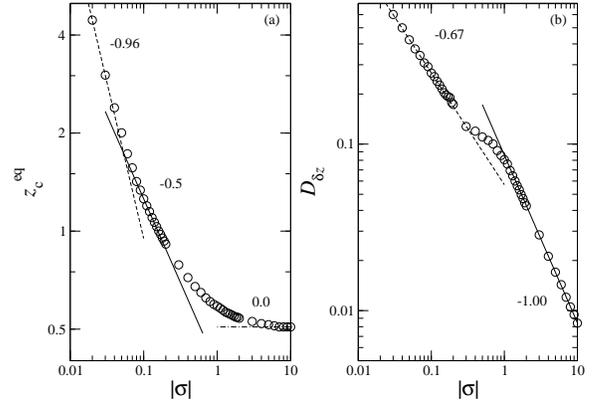}
\end{center}
\vspace{-10pt}
 \caption{(a) Simulation results for the equilibrium height of the
 polymer, $z_{\rm c}^{\rm eq}$ (in units of $2a$), as a function of surface charge density,
 $|\sigma|$ (in units of $q/4a^2$), in logarithmic scale at $\epsilon'/\epsilon=0.01$
 and $\kappa^{-1}=25$.
 It shows the scaling exponents decreases continuously from -0.96
 (weak compression) to 0 (compression) with the surface
 charge density. The scaling exponent of -0.5 indicating the
 crossover (analytically predicted) occurs at $|\sigma|\simeq 0.11$.
 (b) The thickness of the diffusion layer, $D_{\delta z}$, as a function of
 surface charge density, $|\sigma|$, in logarithmic scale. It
 shows the scaling exponents are -0.67 and -1.00 for weak compression and
 compression, respectively. The crossover occurs at $|\sigma|\simeq 0.11$.}
 \label{scal.eps}
%\vspace{-5pt}
\end{figure}

%%%%%%%%%%%%%%%%%%%%%%%%%%%%%%%%%%%%%%%%%%%%%%%%%%%%%%%%%%%%%%%%%%%%%%%%%%%%%%%%%%%%%%%%%

Without the effect from $\vec t(s)$ and $\delta \vec r_\perp(s)$
which expressed in the effective potential of $\delta z(s)$, the
polymer acts as a rigid rod. Its equilibrium height is at $z_{\rm
c}=\sqrt{\Gamma'/4h}$. The ensemble average
\begin{eqnarray} \label{deltaz}
\langle\delta z(s)\rangle = \frac{z_{\rm c}}{2} ( 1- \frac{4h
z_{\rm c}^2}{\Gamma'})
\end{eqnarray}
The entropic force points upward (downward) when the height of
center of mass of the polymer is lower (higher) than
$\sqrt{\Gamma'/4h}$. If $\Gamma'=0$ (same dielectric constants),
the effective potential for $\delta z(s)$ is linear rather than
the harmonic. The result for the case of low-dielectric substrate
cannot be analytically continued to the case of same dielectric
constants. If $\Gamma'<0$ (high dielectric substrate), the system
is unstable. It implies that the decomposition in
Eq.(\ref{new_rs}) is inadequate in high-dielectric case.

%Because of the physically boundary on the surface, the fluctuation
%in $z$-direction cannot be too large, that is, $|\delta z(s)|\ll
%z_{\rm c}$. We approximate the potential up to quadratic order
%only. The effective potential for $\delta z(s)$ is
%\begin{eqnarray}
%V(\delta z) &=& (h-\frac{\Gamma'}{4z_{\rm c}^2})\delta z +
%\frac{\Gamma'}{4z_{\rm c}^3}\delta z^2 \nonumber \\
%&=& \frac{\Gamma'}{4z_{\rm c}^3}(\delta z + \frac{2z_{\rm c
%}^3}{\Gamma'}(h-\frac{\Gamma'}{4z_{\rm c}^2}))^2 - \frac{z_{\rm
%c}^3}{\Gamma'}(h-\frac{\Gamma'}{4z_{\rm c}^2})^2 \nonumber \\
%&\simeq& - \frac{z_{\rm c}^3}{\Gamma'}(h-\frac{\Gamma'}{4z_{\rm
%c}^2})^2
%\end{eqnarray}

Hence, after integrating out the fluctuation variable $\delta
z(s)$ under the ground state dominance (large-$N$ limit), there
leaves only the variable $z_{\rm c}$ in the partition function
which determine the effective probability density distribution for
the height of the center of mass,
\begin{eqnarray} \label{rhozc}
\rho(z_{\rm c})&=&z_{\rm c} \exp[-N\beta ((h z_{\rm
c}+\frac{\Gamma'}{4 z_{\rm c}}) \nonumber \\
&&  + \frac{z_{\rm c}^3}{\Gamma'}(h-\frac{\Gamma'}{4z_{\rm
c}^2})^2 -\frac{a}{2}(\frac{\Gamma'}{2\beta z_{\rm
c}^3})^\frac{1}{2}) ]
\end{eqnarray}
up to a normalization constant. The new equilibrium including the
effect from conformational changes is calculated by ``force
balance", $\partial_{z_{\rm c}}\log\rho(z_{\rm c})=0$, which gives
\begin{eqnarray} \label{zc}
h (z_{\rm c}^{\rm eq})^2 + \frac{a}{4}(\frac{\Gamma'^3}{2\beta
(z_{\rm c}^{\rm eq})^5})^\frac{1}{2} =\frac{\Gamma'}{4}
\end{eqnarray}
at large-$N$ limit. For high enough surface charge density that
$z_{\rm c}^{\rm eq}$ is low, Eq.(\ref{zc}) gives $z_{\rm c}^{\rm
eq}\sim |\sigma|^0$. The polymer is compressed in which the center
of mass is independent of the surface charge density. When the
surface charge density is lowered such that $z_{\rm c}^{\rm eq}$
is high, Eq.(\ref{zc}) reduces to the scaling $z_{\rm c}^{\rm
eq}\sim |\sigma|^{-1/2}$. If the surface charge density is further
lowered so that the polymer basically behaves as a colloid
(undeformed state), Eq.(\ref{rhozc}) becomes $\rho(z_{\rm
c})=\exp[-N\beta hz_{\rm c}]$, and hence $z_{\rm c}^{\rm
eq}\sim|\sigma|^{-1}$. It predicts a continuous crossover from the
compressed state to the weakly compressed state. It is different
from that obtained by scaling analysis for grafted polymer,
predicting a discontinuous jump \cite{borisov}. Our analytical
result is consistent with MC simulation as shown in
Fig.\ref{scal.eps}a.

%%%%%%%%%%%%%%%%%%%%%%%%%%%%%%%%%%%%%%%%%%%%%%%%%%%%%%%%%%%%%%%%%%%%%%%%%%%%%%%%%
\vspace{15pt}
\begin{figure}[tbh]
\begin{center}
\includegraphics[width=3in]{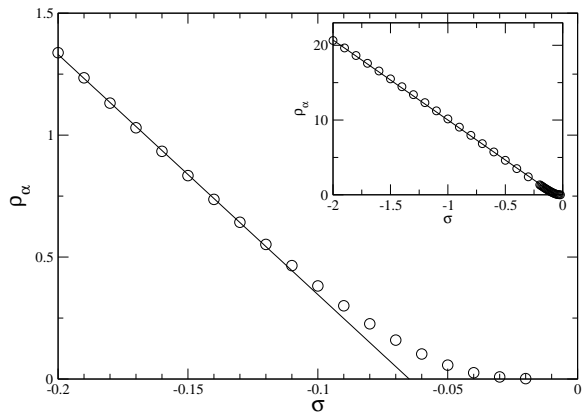}
\end{center}
\vspace{-10pt}
 \caption{Monte-Carlo results for the normalized monomer density at the surface,
 $\rho_a$, as a function of surface charge density, $\sigma$ (in units of $q/4a^2$),
 at $\epsilon'/\epsilon=0.01$ and $\kappa^{-1}=25$.
 The straight line is linearly fit to the data of higher $|\sigma|$.
 It shows the data starts to deviate from the
 linearity when $\sigma \simeq -0.11$. Inset: More results to
 cover a larger range of surface charge density $\sigma$. It shows
 the data follows the linearity at higher $|\sigma|$.
 }
 \label{rho.eps}
\vspace{-5pt}
\end{figure}
%%%%%%%%%%%%%%%%%%%%%%%%%%%%%%%%%%%%%%%%%%%%%%%%%%%%%%%%%%%%%%%%%%%%%%%%%%%%%%%%%%%%%

Besides the position of the center of mass, we also calculate the
thickness of the diffusion layer which is defined as the
characteristic length scale of the exponential decay of monomer
density. We first determine the saddle-point $z_{\rm c}^{*}$ from
Eq.(\ref{part2}) (equivalent to integrating out the variable
$z_{\rm c}$ at large-$N$ limit), which is given by
\begin{eqnarray}
h (z_{\rm c}^*)^3 - \frac{\Gamma'}{4}z_{\rm
c}^*+\Gamma'\langle\delta z(s)\rangle = 0
\end{eqnarray}
The above equation is then solved self-consistently with
Eq.(\ref{deltaz}) in which $z_{\rm c}$ is replaced by $z_{\rm
c}^*$. The solution is
\begin{eqnarray}
z_{\rm c}^*=\sqrt{\Gamma'/4h}\sim |\sigma|^{-\frac{1}{2}}
\end{eqnarray}
The effective partition function for $\delta z(s)$ becomes
\begin{eqnarray}
Z &=& \int{\cal D}[\delta z(s)]
 \exp [\int_0^N ds \{
 -\frac{1}{2a^2}\left(\frac{\partial{\delta
 z(s)}}{\partial{s}}\right)^2 \nonumber \\
&& - \frac{\beta\Gamma'}{4(z_{\rm c}^*)^3}(\delta z(s))^2
  \}]
\end{eqnarray}
which gives the diffusion layer thickness
\begin{eqnarray}
D_{\delta z}\sim (z_{\rm c}^*)^\frac{3}{4}\sim
|\sigma|^{-\frac{3}{8}}
\end{eqnarray}
The scaling exponent does not depend on the surface charge
density. It means that the local fluctuation $\delta z(s)$ is
independent of the polymer conformation, which is consistent with
its definition expressed in Eqs.(\ref{new_rs})-(\ref{new rs1}).
However, simulation results in Fig.\ref{scal.eps}b show that the
scaling exponent is -0.67, a quite large deviation from our
analytical result, -0.375. The deviation may be due to the
approximation of the effective potential up to the quadratic order
only. As shown in Fig.\ref{scal.eps}a and \ref{scal.eps}b, both
the simulation results of $z_{\rm c}^{\rm eq}$ and $D_{\delta z}$
exhibit the crossover between the weak compression and compression
regimes occuring at $\sigma\simeq -0.11$.

Finally we also examine the relation between the surface monomer
density and surface charge density by MC simulation.
Fig.\ref{rho.eps} shows the simulation data follows the linearity
at high enough $|\sigma|$, and start to deviate from the linearity
at $\sigma\simeq -0.11$. The linear relation implies the
compression regime \cite{cheng1}. The deviation from linearity
tells that the polymer starts to be weakly compressed, which is
also consistent with the MC results of both $z_{\rm c}^{\rm eq}$
and $D_{\delta z}$. Since it is hard to characterize the polymer
conformation in MC simulation by the original definition as shown
in Fig.\ref{conf.eps}, we would like to propose an alternative
(operational) definition for charged polymer adsorption - {\it the
linearity between the surface monomer density and the surface
charge density}.

Support by the National Science Council of the Republic of China
is acknowledged under Grant Nos. NSC92-2816-M-008-0005-6, in part
under NSC93-2816-M-001-0007-6 (C.H.C.), and NSC93-2112-M-008-014
(P.Y.L.).

\end{document}